\documentclass[11pt,a4paper]{article}
\usepackage{jheppub} 
\usepackage{graphicx}
\usepackage{amssymb}
\usepackage{amsmath}
\usepackage{subfigure}
\usepackage{float}
\usepackage{bm}
\usepackage[T1]{fontenc} % if needed
\usepackage{array}
\usepackage{hyperref}
\graphicspath{{./figures/}}

\def\simge{\mathrel{%
       \rlap{\raise 0.511ex \hbox{$>$}}{\lower 0.511ex \hbox{$\sim$}}}}
\def\simle{\mathrel{
       \rlap{\raise 0.511ex \hbox{$<$}}{\lower 0.511ex \hbox{$\sim$}}}}

\makeatletter

\newcommand*{\doi}[1]{\href{https://doi.org/#1}{doi:\ #1}}
\newcommand{\figcaption}[1]{\def\@captype{figure}\caption{#1}}
\newcommand{\tblcaption}[1]{\def\@captype{table}\caption{#1}}

\newcommand{\Nsite}{N_{\rm full}}
\newcommand{\Nsub}{N_{\rm sub}}

\newcommand{\qfull}{Q_{\rm full}}
\newcommand{\qsub}{Q_{\rm sub}}

\newcommand{\nape}{n_{\rm APE}}

\newcommand{\SU}{$\mathrm{SU}$}

\renewcommand{\th}{\theta}

\makeatother

\begin{document}
\title{Subvolume method for \SU(2) Yang-Mills theory at finite
temperature: topological charge distributions}

\def\KEK{High Energy Accelerator Research Organization (KEK), Tsukuba 305-0801, Japan}
\def\SOKENDAI{Graduate University for Advanced Studies (SOKENDAI), %
Tsukuba 305-0801, Japan}
\def\IPMU{Kavli Institute for the Physics and Mathematics of the
Universe (WPI), University of Tokyo, Kashiwa, Chiba 277-8583, Japan}
\def\CDDD{Center for Data-Driven Discovery, Kavli IPMU (WPI), 
University of Tokyo, Kashiwa, Chiba 277-8583, Japan}
\def\TSQS{Trans-Scale Quantum Science Institute, The University of Tokyo, Tokyo 113-0033, Japan}

\author[a,b]{Norikazu Yamada,}
\emailAdd{norikazu.yamada@kek.jp}
\author[c,d,e]{Masahito Yamazaki,}
\emailAdd{masahito.yamazaki@ipmu.jp}
\author[a,b]{Ryuichiro Kitano}
\emailAdd{ryuichiro.kitano@kek.jp}
\affiliation[a]{\KEK}
\affiliation[b]{\SOKENDAI}
\affiliation[c]{\IPMU}
\affiliation[d]{\CDDD}
\affiliation[e]{\TSQS}

\date{\today}

\abstract{
 We apply the previously-developed sub-volume method to study the
 $\th$-dependence of the four-dimensional \SU(2) Yang-Mills theory
 at finite temperature.
 We calculate the first two coefficients, the topological susceptibility
 $\chi$ and the fourth cumulant $b_2$, in the $\th$-expansion of the
 free energy density around the critical temperature ($T_c$) for the
 confinement-deconfinement transition.
 Lattice calculations are performed with three different spatial sizes
 $24^3,32^3,48^3$ to monitor finite size effects, while the temporal
 size is fixed to be $8$.
 The systematic uncertainty associated with the sub-volume extrapolation
 is studied with special care.
 The sub-volume method allows us to determine the values of $b_2$ much
 more accurately than the standard full-volume method, and we successfully
 identify the temperature dependence of $b_2$ around $T_c$.
 Our numerical results suggest that the $\th$-dependence of the free
 energy density near $\th=0$ changes from $4\chi(1-\cos(\th/2))$ to
 $\chi(1-\cos\th)$ as the temperature crosses $T_c$.
}

\maketitle

%%%%%%%%%%%%%%%%%%%%%%%%%%%%%%%%%%%%%%%%%
\section{Introduction}\label{sec:intrroduction}

It has been a long-standing problem to identify the effect of the
$\theta$ parameter \cite{Callan:1976je} on the dynamics of
four-dimensional pure \SU($N_c$) Yang-Mills (YM) theory.
The problem can be further enriched by considering the theory at finite
temperature $T$.

While there have been attempts to study the $\theta$--$T$ phase diagram,
the previous studies were restricted to limited regions of it.
Along the temperature axis ({\it i.e.} $\theta=0$), the existence of the
confinement-deconfinement phase transition has been well established,
and the critical temperature ($T_c$) and other properties of the phase
transition were explicitly determined on the lattice for some 
theories~\cite{Lucini:2002ku,Lucini:2003zr}.
It is also possible to explore the small-$\theta$ region by various
techniques, such as reweighting, Taylor expansion in $\theta$, or
analytic continuation into imaginary
$\th$~\cite{DElia:2012pvq,DElia:2013uaf,Bonati:2013tt,Bonati:2016tvi,Kitano:2020mfk,Otake:2022bcq,Borsanyi:2022fub,Bonanno:2023hhp,Hirasawa:2024vns,Bonanno:2024ggk}.

Recently a novel analysis method, called \textit{the sub-volume method},
was developed in~\cite{Kitano:2021jho} to investigate the region $\th\sim O(1)$
for the pure \SU(2) YM theory at zero and finite temperatures\footnote{
The $\th\sim O(1)$ region is explored also in~\cite{Hirasawa:2024vns} relying
on analytic continuation.}.
The result suggests spontaneous CP violation at $\theta=\pi$ in
the vacuum and the restoration above $T_c$, thus qualitatively the same
picture as the large $N_c$
limit~\cite{tHooft:1973alw,Witten:1980sp,tHooft:1981bkw,Witten:1998uka}
emerges; see
also~\cite{Unsal:2012zj,Gaiotto:2017yup,Kitano:2017jng,Yamazaki:2017ulc,Yamazaki:2017dra,Nomura:2017zqj,Wan:2018zql}
for samples of related analysis for finite $N_c$ pure YM theory.

In this paper we study the $\theta$ dependence of the free energy
density of the pure \SU(2) YM theory near the critical temperature,
where the free energy density is expected to undergo a qualitative
change.
We determine the first two coefficients appearing in the Taylor
expansion of the free energy density, the topological susceptibility
$\chi$ and the fourth cumulant $b_2$.
While the lattice determinations of these quantities across $T_c$ are
available for $N_c \ge 3$ in~\cite{Bonati:2013tt,Borsanyi:2022fub},
little has been done in $N_c=2$ except for~\cite{Lucini:2004yh}.
Furthermore, the determinations of $\chi$ and especially $b_2$ require high
statistics.
Indeed, $O(100,000)$ configurations had to be accumulated to reach a
reasonable accuracy for the calculation of $b_2$ in \SU($2$) YM theory
at $T=0$~\cite{Kitano:2020mfk}.
The situation, however, is significantly improved in the newly-proposed
sub-volume method~\cite{Kitano:2021jho} as described below.
The basic idea of the sub-volume method is inspired by the work of~\cite{KeithHynes:2008rw}
done in two dimensions.
The slab
method~\cite{deForcrand:1998ng,Bietenholz:2015rsa,Bietenholz:2016szu}
and the method using the sub-volume introduced in~\cite{LSD:2014yyp} were proposed
to resolve the problem of frozen topology and are conceptually different from but
operationally similar to ours.

The rest of this paper is organized as follows.
The lattice setup and parameters are summarized in sec.~\ref{sec:setup},
which also includes the definition of the local topological charge on
the lattice as well as the explanation of the sub-volume method.
In sec.~\ref{sec:results}, we present our main results and discuss the
detailed numerical analysis step by step.
The sub-volume method has a subtlety in the estimate of uncertainties
associated with the large sub-volume extrapolations.
Section~\ref{sec:sys-error} is devoted to the discussion of this
uncertainty.
Motivated by the numerical results obtained here, we conclude that $b_2$
is approximately independent of $T$ in the confined phase and infer the whole
$\th$-dependence of the free energy density below $T_c$ in
sec.~\ref{sec:discussion}.
Finally, the study is summarized in sec.~\ref{sec:summary}.

%%%%%%%%%%%%%%%%%%%%%%%%%%%%%%%%%%%%%%%%%
\section{Lattice set-up and methods}\label{sec:setup}

\subsection{Lattice parameters}\label{subsec:parameters}

The \SU(2) lattice gauge action at $\theta=0$ is described by
\begin{align}
 S_g = 6\,\beta\,\Nsite\,\left\{c_0(1-W_P)+2\,c_1(1-W_R)\right\}
 \ ,
 \label{eq:action}
\end{align}
where $\beta=4/g^2$ is the inverse lattice gauge coupling.
The $1\times 1$ plaquette and the $1\times 2$ rectangle are constructed
from \SU(2) link variables, and $W_P$ and $W_R$ are those averaged over
four dimensional lattice sites ($\Nsite=N_S^3\times N_T$) and all
possible directions, respectively.
The coefficients $c_0$ and $c_1$ satisfying $c_0=1-8c_1$ are the improvement
coefficients, and we take the tree-level Symanzik improved action,
$c_1=-1/12$~\cite{Weisz:1982zw}.
To monitor the stability of numerical results, the calculations are
carried out on three lattice sizes ($N_S=$24, 32, 48) with $N_T=8$
fixed.
The critical value of $\beta$ at $\theta=0$ for $N_T=8$, corresponding to the
critical temperature $T_c$, is known to be
$\beta_c=1.920(5)$~\cite{Cella:1993ic,Giudice:2017dor}, around which we perform
the numerical simulations.

Gauge configurations are generated by the Hybrid Monte Carlo method with
periodic boundary conditions in all directions, and stored in every ten
trajectories after thermalization.
Statistical errors are estimated by the single-elimination jack-knife
method with the bin size of 1,000 configurations.
Simulation parameters including the lattice size and the statistics are
summarized in Tab.~\ref{tab:parameters}.
%%%%%%%%%%%%%%%%%%%%%%%%%%%%%%%%%%%%%%%%
\begin{table}[tbp]
 \begin{center}
 \begin{tabular}{|p{6ex}| p{4ex}p{4ex}p{4ex}p{4ex}p{4ex}p{4ex}p{4ex}p{4ex}p{4ex}p{4ex}p{4ex}p{4ex}p{4ex}|}
 \hline
  $\beta$ & 1.80 & 1.85 & 1.86 & 1.87 & 1.88 & 1.89 & 1.90 & 1.91 & 1.92 & 1.93 & 1.94 & 1.95 & 2.00\\
  $T/T_c$ & 0.68 & 0.80 & 0.83 & 0.85 & 0.88 & 0.91 & 0.94 & 0.97 & 1.00 & 1.03 &      & 1.10 & 1.30\\
 \hline
  $\!\!N_S$=24~
          &  &  &30&30&20&20&20&32&50&20&  &  &\\
        32&30&30&10&20&40&30&30&54&50&50&50&31&10\\
        48&25&  &  &10&  &20&20&30&38&12&  &12&10\\
  \hline
 \end{tabular}
 \caption{
  The simulation parameters and statistics.
  The numbers of configurations are shown in units of 1,000.
  The temporal size $N_T$ is fixed to 8.
  The normalized temperature $T/T_c$ is obtained by fitting the data presented in~\cite{Giudice:2017dor}.
  }
 \label{tab:parameters}
 \end{center}
\end{table}
%%%%%%%%%%%%%%%%%%%%%%%%%%%%%%%%%%%%%%%%

\subsection{Topological charge density on the lattice}
\label{subsec:topological-charge-density}

The sub-volume method, described in the next subsection, requires the
local values of the topological charge density.
The conceptually cleanest way to measure them would be to use the
eigenvectors of the overlap Dirac operator~\cite{Hasenfratz:1998ri}.
However, this method is not realistic for high-statistics and
large-volume lattice calculations because the computational cost is too
large.
Instead, we use the bosonic definition, where 
the topological term $G\tilde{G}$ is directly
constructed from the \SU(2) link variables.
The topological charge distribution thus obtained is distorted by
lattice artifacts, which need to be eliminated by the smearing procedure.
The smearing, however, may simultaneously deform physically legitimate
topological excitations.
Based on the study of this point in~\cite{Kitano:2020mfk}, 
we estimate the physical topological quantities by extrapolating the
smeared data over a suitable range of the smearing steps to the zero smearing.

Among several mutually-consistent methods often used in the
literature~\cite{Bonati:2014tqa,Alexandrou:2015yba,Alexandrou:2017hqw},
we choose the APE smearing~\cite{Albanese:1987ds} and the five-loop
improved topological charge operator~\cite{deForcrand:1997esx} to calculate
the local topological charge $q(x,\nape)$ on each configuration.
This procedure and the parameters for the APE smearings are precisely the same
as those used in~\cite{Kitano:2020mfk,Kitano:2021jho}.

A local topological charge is calculated at every five smearing steps,
{\it i.e.} $n_{\rm APE}=$ 0, 5, 10, $\cdots, 50$ and is uniformly
shifted as $q(x,\nape)\to q(x,\nape)+\epsilon$ such that the global
topological charge $Q_{\rm full}:=\sum_{x\in \Nsite}q(x,\nape)$ takes
the integer value closest to the original one.

\begin{figure}[tb]
  \begin{center}
  \begin{tabular}{cc}
  \includegraphics[width=0.5 \textwidth, trim=30 0 0 0]{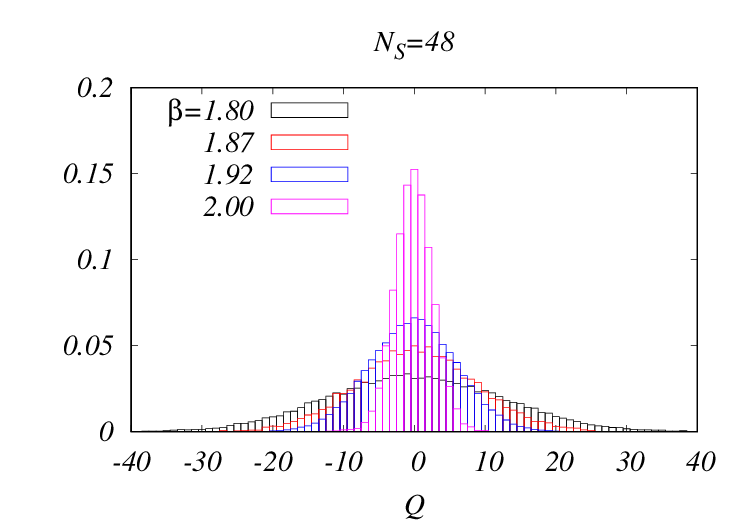} &
  \includegraphics[width=0.5 \textwidth, trim=30 0 0 0]{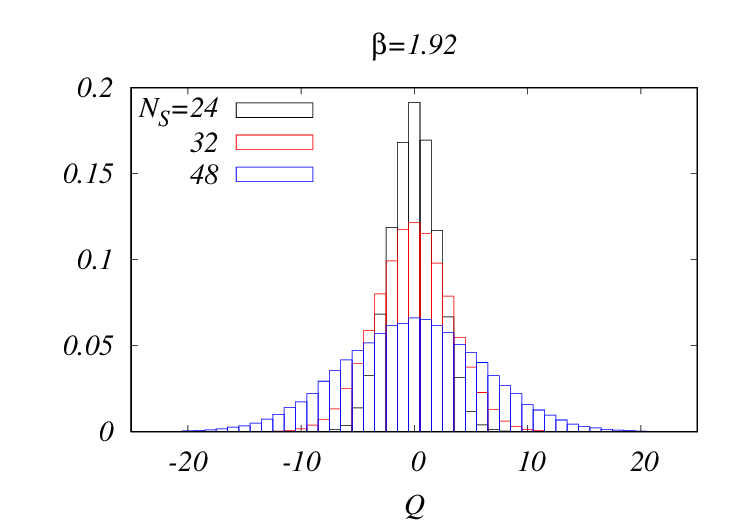}
  \end{tabular}
  \end{center}
 \caption{Histogram of global topological charge, $Q_{\rm full}$, at
 $\nape=30$ at various temperatures (or $\beta$) on $N_S=48$ lattice
 (left) and at $T_c$ (or $\beta_c$) on three different $N_S$ lattices
 (right).
 }
 \label{fig:Q-histogram}
\end{figure}

Figure~\ref{fig:Q-histogram} shows the histograms of global topological
charge $Q_{\rm full}$ thus obtained at $\nape=30$.
It is seen that topology fluctuates appropriately for all the values of $\beta$
and $N_S$ studied in this paper.
%%%%%%%%%%%%%%%%%%%%%%%%%%%%%%%%%%%%%%%%
%%%%%%%%%%%%%%%%%%%%%%%%%%%%%%%%%%%%%%%%

%%%%%%%%%%%%%%%%%%%%%%%%%%%%%%%%%%%%%%%%%%%%
\subsection{Sub-volume method}
\label{sec:method}

To study of the $\th$-dependence of the free energy density $f(\theta)$,
we consider the Taylor expansion around $\th=0$,
\begin{align}
 f(\th)=\frac{\chi\th^2}{2}(1+b_2 \th^2+\cdots)\,.
 \label{eq:Taylor-e}
\end{align}
In the sub-volume method, the first two coefficients $\chi$ and $b_2$ are obtained as
\begin{align}
&  \chi_{\rm sub} = \lim_{\nape\to 0}\lim_{l\to \infty} \chi_{\rm sub}(\nape,l)
= \lim_{\nape\to 0}\lim_{l\to \infty}\frac{1}{\Nsub}
  \left\langle Q_{\rm sub}^2(\nape) \right\rangle\,,
 \label{eq:chi}\\
&  b_{2,\rm sub} = \lim_{\nape\to 0}\lim_{l\to \infty} b_{2,\rm sub}(\nape,l)
= \lim_{\nape\to 0}\lim_{l\to \infty}
  \frac{-  \left\langle Q_{\rm sub}^4(\nape) \right\rangle
        + 3\left\langle Q_{\rm sub}^2(\nape) \right\rangle^2}
       { 12\left\langle Q_{\rm sub}^2(\nape) \right\rangle}\,,
 \label{eq:b2}
\end{align}
where 
\begin{align}
& \qsub(\nape) := \displaystyle\sum_{x\in \Nsub} q(x,n_{\rm APE})\ .
\end{align}
and the sub-volume sizes are $\Nsub=l^3\times 8$ with
$l=8, 12, 16, \cdots , N_S$.
We have taken full advantage of the translational invariance and hence the results from
smaller sub-volumes are statistically more accurate.
While we can exchange the order of the $\nape\to 0$ limit and the large
sub-volume limit $l\to \infty$, we have chosen the ordering as above
which gives slightly better statistical uncertainties.

As pointed out in~\cite{Kitano:2021jho}, the sizes of sub-volume (or $l$)
used in the large sub-volume extrapolation have to be carefully chosen,
and the suitable range is not known in advance.
A reasonable choice for the lower end of the range would be
$l_{\rm min}=N_T$, which corresponds to $1/T$ in physical unit.
On the other hand, the upper end has to be chosen such that the
sub-volume does not see the boundary of the lattice.
It seems reasonable to set $l_{\rm max}\le N_S/2$.
We will decide $l_{\rm max}$ by comparing the data from three lattices
with different $N_S$'s.

In the above expression, replacing $\qsub$ and $\Nsub$ with $\qfull$ and
$\Nsite$ respectively ends up with the standard method using the full-volume data.
We denote the resulting values of $\chi$ and $b_2$ by $\chi_{\rm full}$ and
$b_{2,\rm full}$, respectively.

%%%%%%%%%%%%%%%%%%%%%%%%%%%%%%%%%%%%%%
\section{Numerical results}\label{sec:results}
\subsection{Sub-volume extrapolation}\label{subsec:volume-extrapolation}

\begin{figure}[tb]
  \begin{center}
  \begin{tabular}{cc}
  \includegraphics[width=0.5 \textwidth, trim=30 0 0 0]{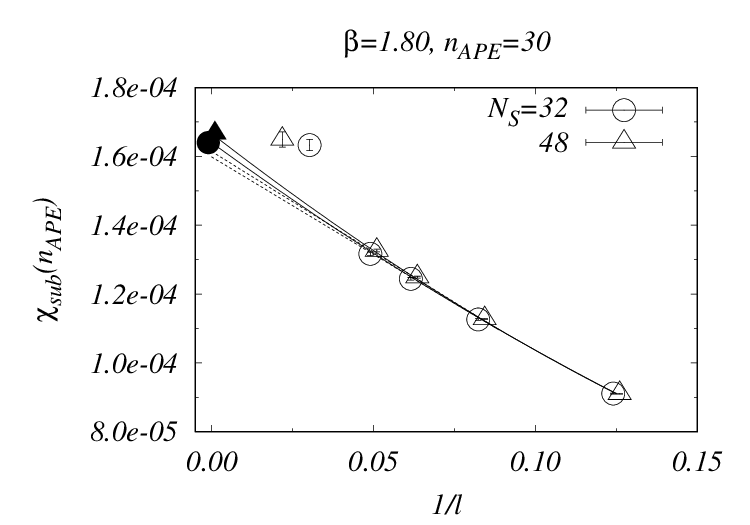}&
  \includegraphics[width=0.5 \textwidth, trim=30 0 0 0]{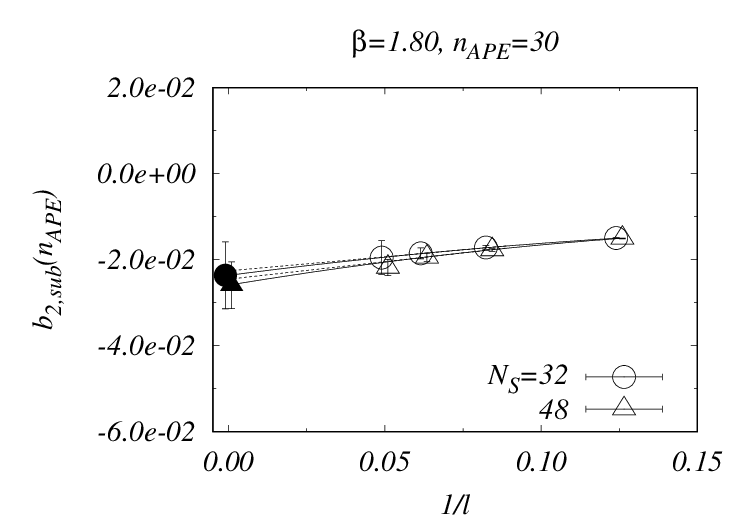}\\
  \includegraphics[width=0.5 \textwidth, trim=30 0 0 0]{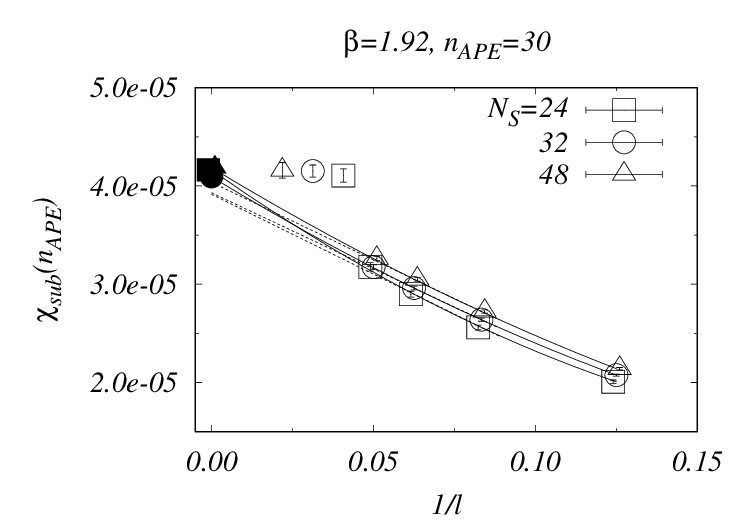}&
  \includegraphics[width=0.5 \textwidth, trim=30 0 0 0]{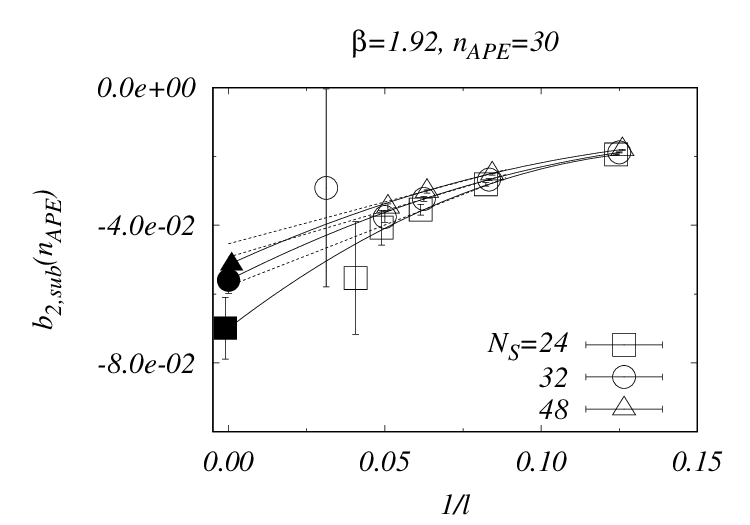}\\
  \includegraphics[width=0.5 \textwidth, trim=30 0 0 0]{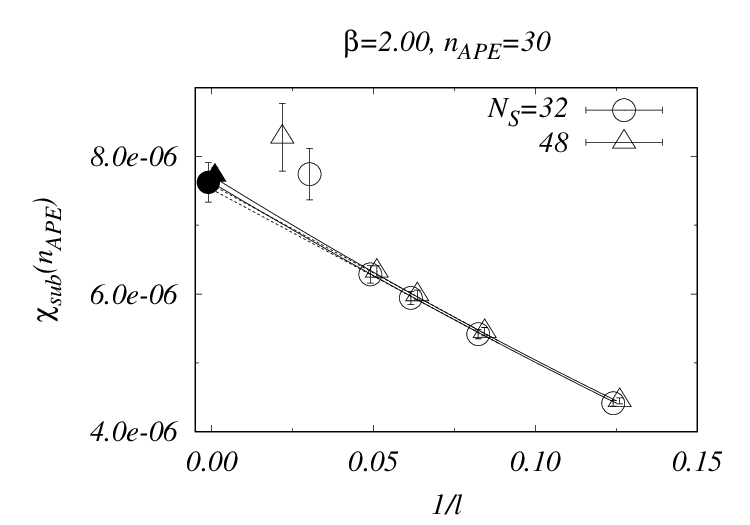}&
  \includegraphics[width=0.5 \textwidth, trim=30 0 0 0]{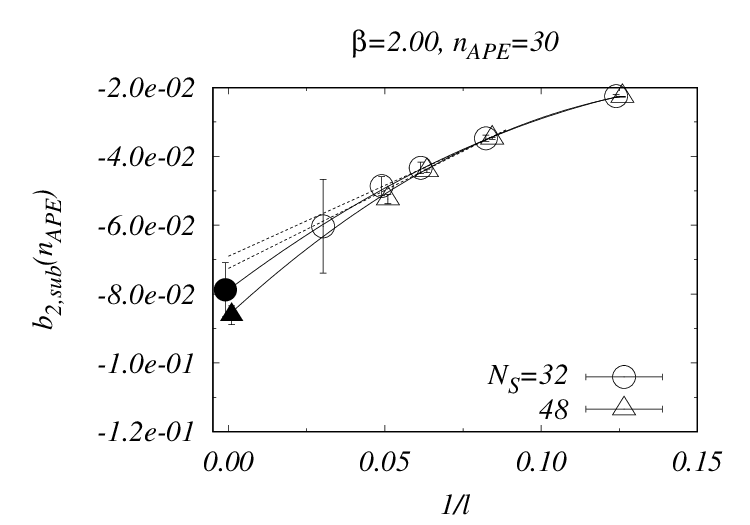}
  \end{tabular}
  \end{center}
 \caption{
 The sub-volume dependence of $\chi_{\rm sub}$ and $b_{2,\rm sub}$.
 The quadratic (solid) and linear (dotted) extrapolations to the large
 sub-volume limit are shown.
 }
 \label{fig:v-vs-chi-b2}
\end{figure}

We calculate $\chi_{\rm sub}$ and $b_{2,\rm sub}$ following
\eqref{eq:chi} and \eqref{eq:b2}, respectively.
Figure~\ref{fig:v-vs-chi-b2} shows the sub-volume (precisely $1/l$) dependence
of $\chi_{\rm sub}(\nape,l)$ and $b_{2,\rm sub}(\nape,l)$, obtained for
various ensembles.
The data at $\nape=30$ are shown as an example.
%%%%%%%%%%%%%%%%%%%%%%%%%%%%%%%%%%%%%%%%
%%%%%%%%%%%%%%%%%%%%%%%%%%%%%%%%%%%%%%%%
It is seen that for both quantities the data obtained with different $N_S$ are well
consistent with each other within uncertainties as long as
$l \simle N_S/2$.
Moreover, the statistical errors of
$b_{2,\rm sub}$ grow more rapidly than those of $\chi_{\rm sub}$ as $l$
approaches $N_S$.

As a general tendency, the deviation among different $N_S$'s starts to
appear in the data on smaller lattice.
This is because the sub-volume starts to be affected by the boundary of
the lattice when $l \simge N_S/2$.
Note that when $l=N_S$ the sub-volume method is nothing but the full volume
method under the periodic boundary condition.

In either plots, there is a range of $l$ for which the data on the two
largest $N_S$ lattices are consistent.
We choose the fit range in the large-sub-volume extrapolation to be
$l=[8, 16]$ for all cases, and omit the results with $N_S=24$ from the
main analysis.

Within the fit range, the data are not aligned in a straight line
especially for $b_{2,\rm sub}$, and hence we chose to extrapolate by
a quadratic function in $1/l$.
To estimate the size of the systematic uncertainty associated
with the extrapolation, we also performed a linear extrapolation using
two data points at $l=12$ and 16.
The results of the extrapolations with quadratic and linear functions are
shown by solid curves and dotted lines in Figs.~\ref{fig:v-vs-chi-b2}, respectively.

\subsection{Extrapolation to \texorpdfstring{$\nape=0$}{nAPE=0}}
  \label{subsec:nape-extrapolation}

\begin{figure}[t]
 \begin{center}
  \begin{tabular}{cc}
   \includegraphics[width=0.5 \textwidth, trim=30 0 0 0]{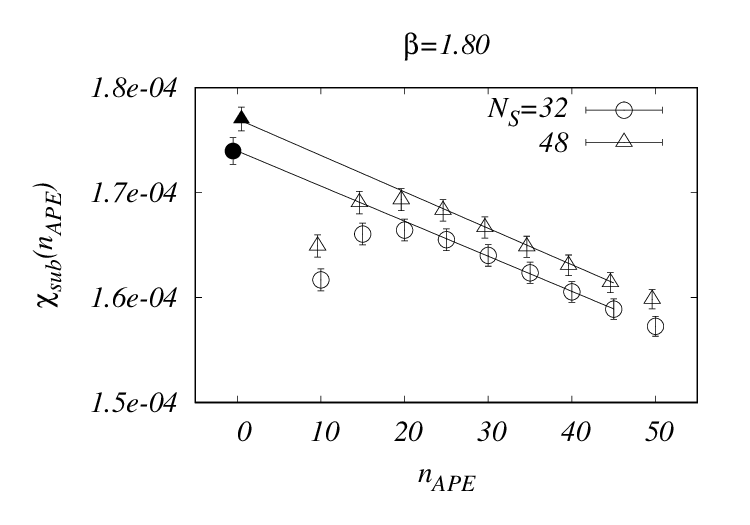}&
   \includegraphics[width=0.5 \textwidth, trim=30 0 0 0]{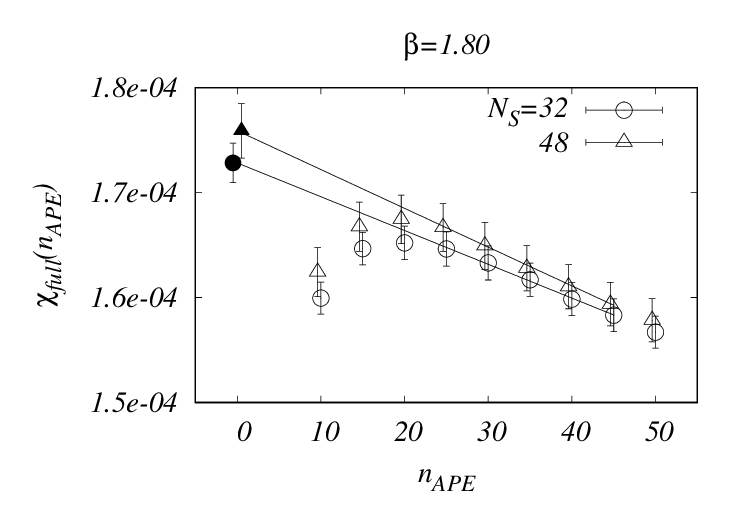}\\
   \includegraphics[width=0.5 \textwidth, trim=30 0 0 0]{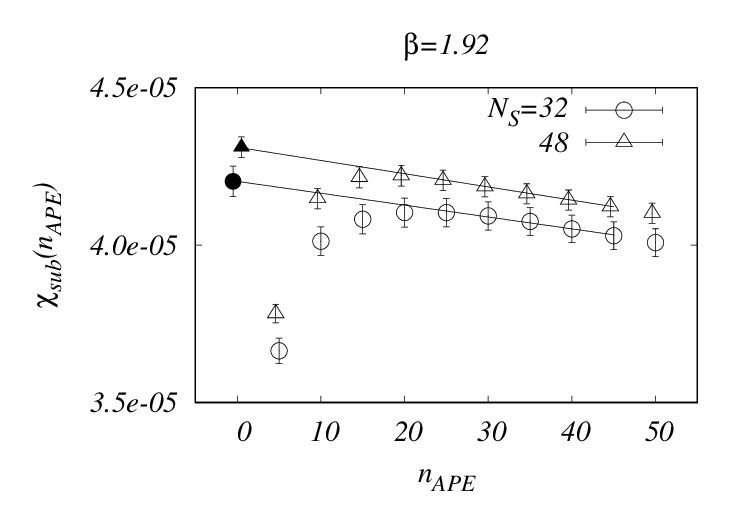}&
   \includegraphics[width=0.5 \textwidth, trim=30 0 0 0]{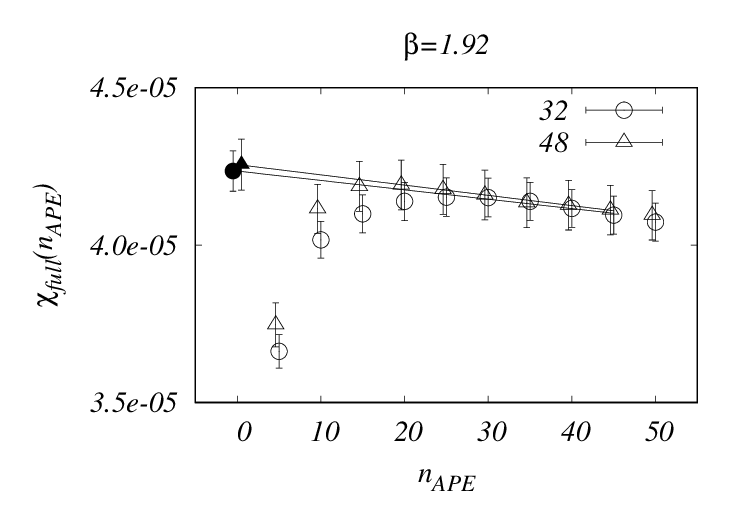}\\
   \includegraphics[width=0.5 \textwidth, trim=30 0 0 0]{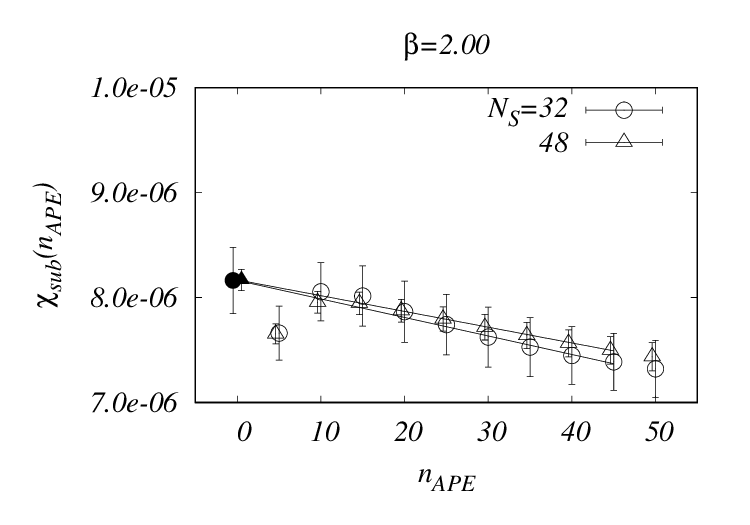}&
   \includegraphics[width=0.5 \textwidth, trim=30 0 0 0]{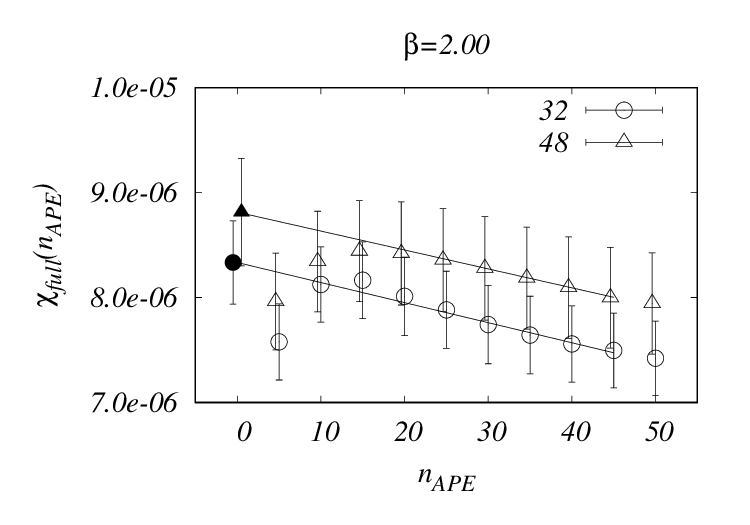}\\
  \end{tabular}
 \end{center}
 \caption{
 The $\nape$ dependence of $\chi$ obtained with sub- (left) and
 full-volume (right) methods.
 Linear extrapolations to $\nape =0$ are also shown (solid lines).
 }
 \label{fig:nape-chi}
\end{figure}

\begin{figure}[t]
 \begin{center}
  \begin{tabular}{cc}
   \includegraphics[width=0.5 \textwidth, trim=30 0 0 0]{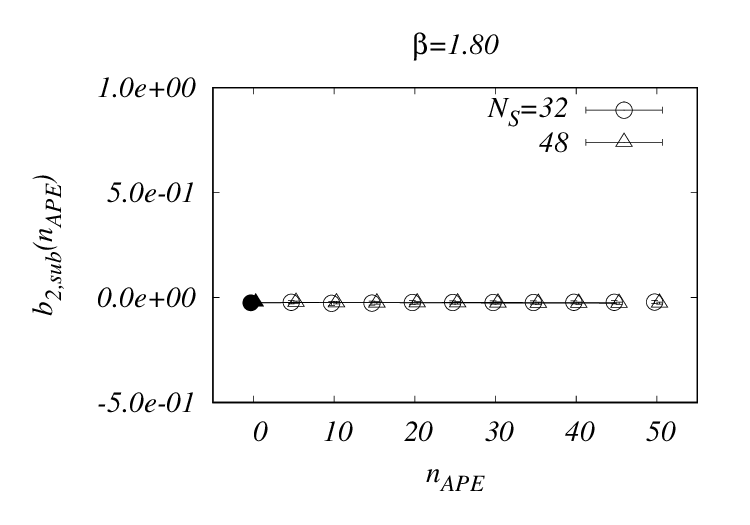}&
   \includegraphics[width=0.5 \textwidth, trim=30 0 0 0]{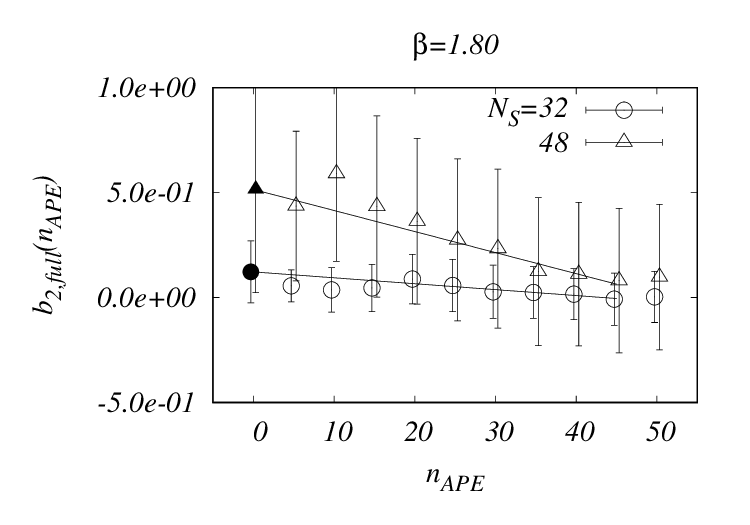}\\
   \includegraphics[width=0.5 \textwidth, trim=30 0 0 0]{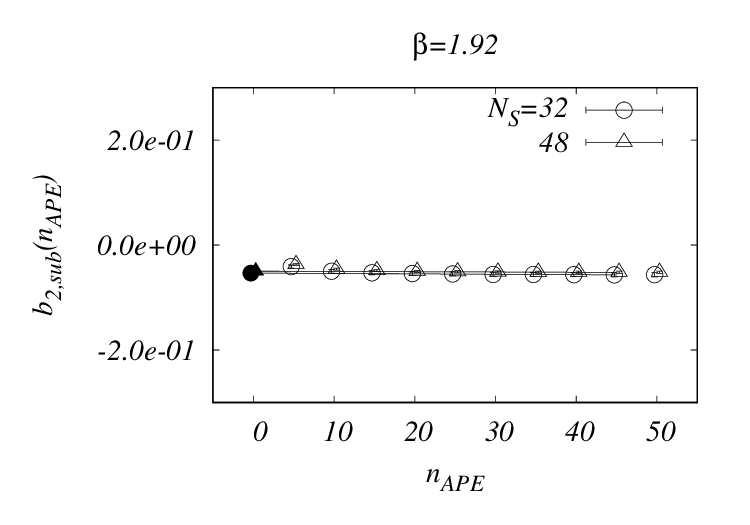}&
   \includegraphics[width=0.5 \textwidth, trim=30 0 0 0]{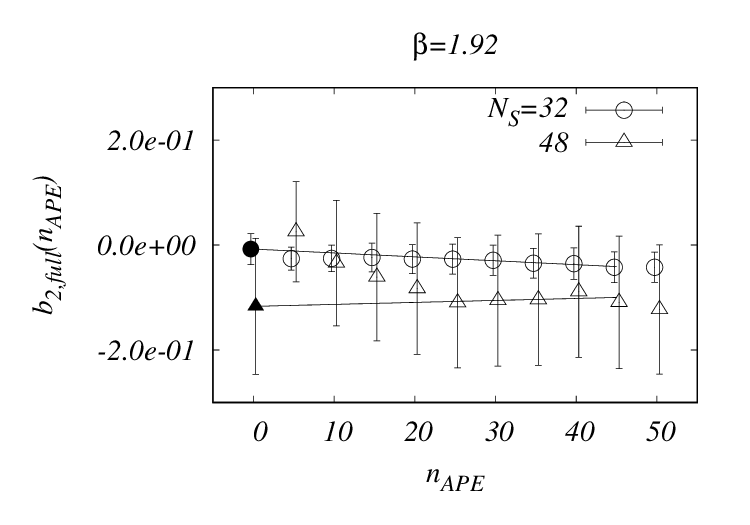}\\
   \includegraphics[width=0.5 \textwidth, trim=30 0 0 0]{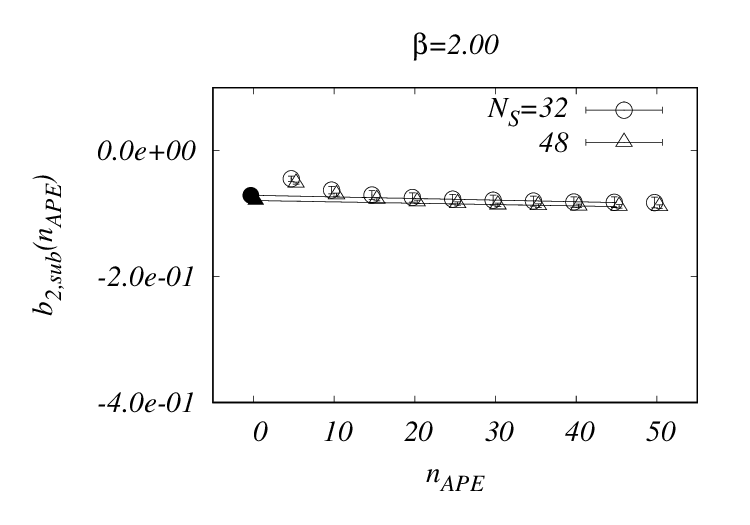}&
   \includegraphics[width=0.5 \textwidth, trim=30 0 0 0]{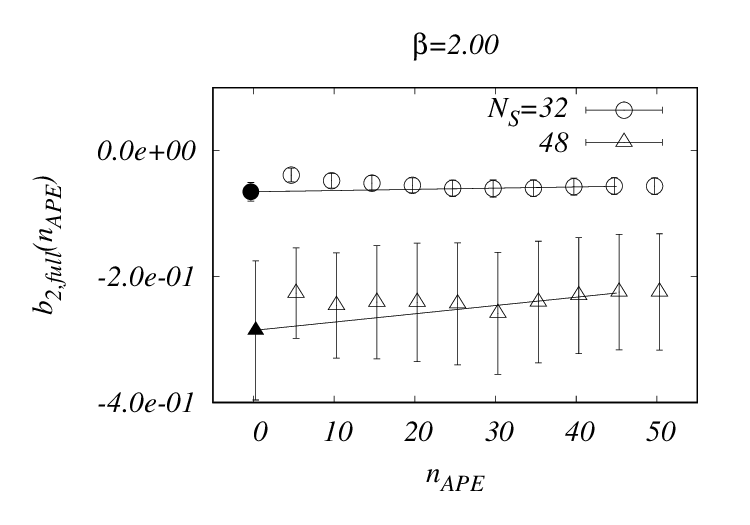}
  \end{tabular}
 \end{center}
 \caption{
 The $\nape$ dependence of $b_2$ obtained with sub- (left) and
 full-volume (right) methods.
 Linear extrapolations to $\nape =0$ are also shown (solid lines).
 }
 \label{fig:nape-b2}
\end{figure}

Next, the data in the large-sub-volume limit obtained at various values of $\nape$
are collected and extrapolated to $\nape=0$.
In Figs.~\ref{fig:nape-chi} and \ref{fig:nape-b2}, the extrapolations of
$\chi$ and $b_2$ are shown respectively, where the data from the sub-
(left) and the full-volume (right) methods are compared.
%%%%%%%%%%%%%%%%%%%%%%%%%%%%%%%%%%%%%%%%
%%%%%%%%%%%%%%%%%%%%%%%%%%%%%%%%%%%%%%%%
%%%%%%%%%%%%%%%%%%%%%%%%%%%%%%%%%%%%%%%%
%%%%%%%%%%%%%%%%%%%%%%%%%%%%%%%%%%%%%%%%
In the plots for the sub-volume calculations (left), we only show the
results from the quadratic extrapolation in $1/l$ for clarity.
The data plotted in the full-volume calculations (right) are those
obtained at $l=N_S$ without any volume extrapolation.
Since $\nape$ dependence is mild, the extrapolation is performed
linearly with the fit range of $\nape=[25, 45]$ in all cases.
A different choice of the fit range, for example $\nape=[20,40]$,
gives only negligible difference.

We choose the same vertical scales on the left and right plots.
By comparing the those plots, we find that the statistical uncertainty
in the sub-volume method is slightly smaller for $\chi$ and
significantly smaller for $b_2$ than those from the full-volume method.

\subsection{Main results}\label{subsec:main-results}

\begin{figure}[tb]
 \begin{center}
  \begin{tabular}{cc}
   \includegraphics[width=0.5 \textwidth, trim=30 0 0 0]{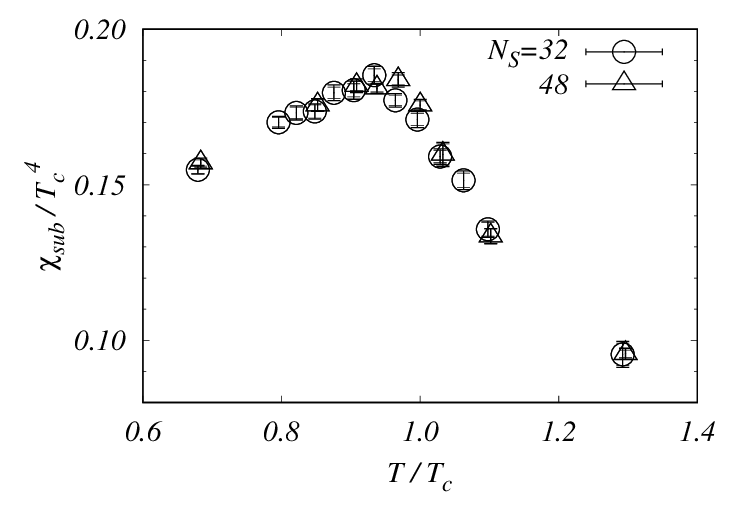}&
   \includegraphics[width=0.5 \textwidth, trim=30 0 0 0]{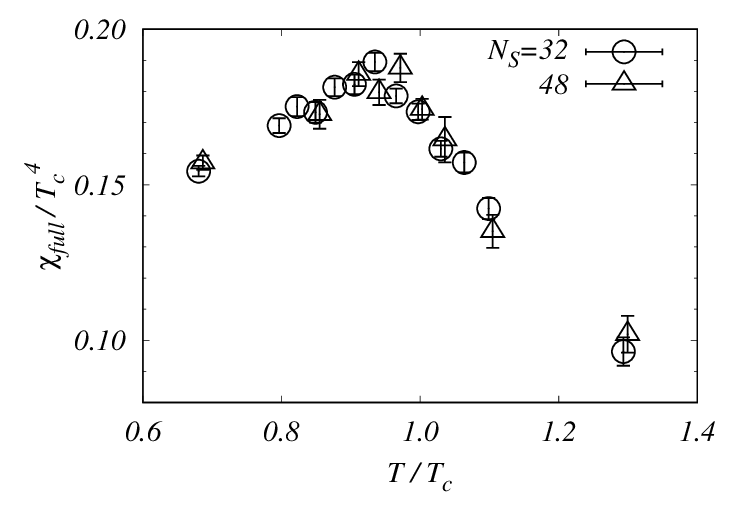}\\
   \includegraphics[width=0.5 \textwidth, trim=30 0 0 0]{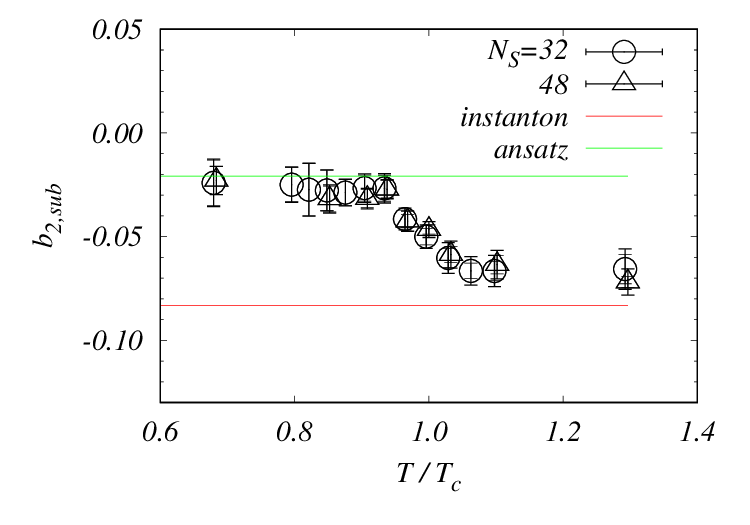}&
   \includegraphics[width=0.5 \textwidth, trim=30 0 0 0]{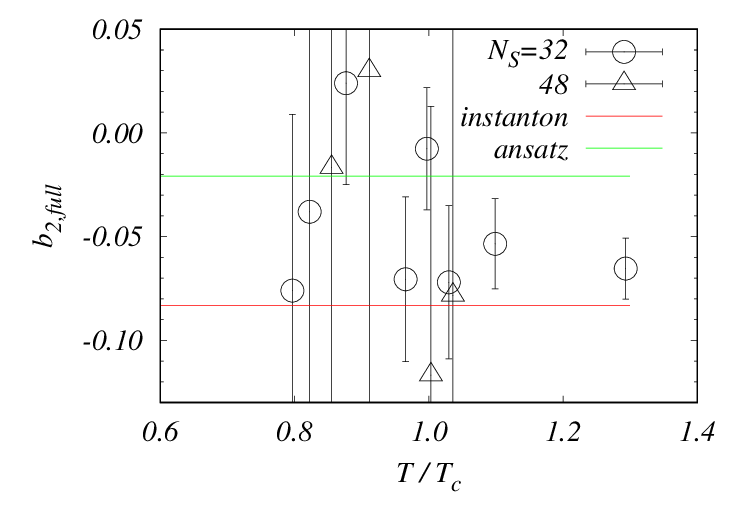}
  \end{tabular}
 \end{center}
 \caption{Temperature dependence of $\chi/T_c^4$ and $b_2$ with the
 sub-volume  (left) and full-volume (right) method.
% The result at $T=0$ are also shown on the left.
 The instanton prediction $b_2=-1/12$ and a phenomenological ansatz
 $b_2=-1/48$ are shown by horizontal lines.
 }
 \label{fig:bdep-chi-b2}
\end{figure}

After two limiting procedures, we obtain $\chi$ and $b_2$ over the
values of $\beta$ and $N_S$ we have explored.
We now try to identify the temperature dependence of these quantities
obtained in the sub- and full-volume methods.
The $T$-dependence of $\chi/T_c^4$ and $b_2$ for $N_S=32$ and 48 are
shown in Fig.~\ref{fig:bdep-chi-b2}.
In the plot, the central value at each point is defined by the average of
the results with the quadratic and linear extrapolations in $1/l$, and
the half of their difference is assigned to the systematic
uncertainty\footnote{For further discussion on the systematic
uncertainty, see the next subsection.}.
The inner and outer error bars in the sub-volume results represent the
statistical error and the sum of the statistical and systematic ones,
respectively.
%%%%%%%%%%%%%%%%%%%%%%%%%%%%%%%%%%%%%%%%
%%%%%%%%%%%%%%%%%%%%%%%%%%%%%%%%%%%%%%%%
It is seen that the systematic error for $b_{2,\rm sub}$ is
relatively large at higher $T$, which reflects a large curvature in the
$1/l$ dependence of $b_{2,\rm sub}$ at such temperatures.

For the $T$-dependence of $\chi$, the results from the sub- and
full-volume are in good agreement, and those with $N_S=32$ and 48 are
also mutually consistent.
We hence conclude that the finite size effects are under control.
These consistency strongly supports the validity of our procedure
of the large-sub-volume extrapolation including the choice of the
functional form and the fit range.

In the \SU(3) case, $\chi$ is almost a constant below $T_c$ and
experiences a gap at $T_c$ before steadily decreasing above $T_c$
to follow the instanton prediction (see, for
example,~\cite{DElia:2012pvq,DElia:2013uaf,Frison:2016vuc,Otake:2022bcq,Borsanyi:2022fub,Bonanno:2023hhp}).
In the \SU(2) case, a mild peak is observed slightly below $T_c$, and no
gap exists at $T=T_c$.
We suspect that this nontrivial $T$-dependence originates from the
temperature dependence of the glueball mass~\cite{Caselle:2013qpa}.
While at $T=0$ the lightest glueball is known to have a mass much larger
than the dynamical scale $\Lambda \sim T_c$, it needs to be massless at
$T=T_c$ since the phase transition is second-order.
Thus it seems reasonable to imagine that at slightly below $T_c$ light
glueballs contribute to the free energy, thereby yielding a nontrivial
$T$-dependence of $\chi$.
Such a temperature-dependence of the glueball mass has not been reported
in the literature, and hence it is interesting to study the dependence
and compare the \SU(2) case with other \SU($N_c$) theories.

For the subleading coefficient $b_2$, the statistical error is improved
remarkably in the sub-volume method, and this has led to a clear
demonstration of its temperature dependence.
The data above $T_c$ are close to 
the value $b_2=-1/12$ (shown as one of the horizontal lines) as
predicted by the dilute instanton calculus \cite{tHooft:1976snw}, and we
expect $b_2$ to further approach the instanton prediction as one goes to
higher temperature.
Below $0.95\,T_c$, $b_2$ again appears to be $T$-independent and
consistent with the value $-1/48$, which is shown as another horizontal
line. We discuss the possible origin of this value in the next section.

The $T$-dependence of $b_2$ is qualitatively similar to the
\SU(3) case except around the transition region, where a negative peak
is observed in \SU(3)~\cite{Borsanyi:2022fub}.
It is interesting to see how the different $T$-dependence of $b_2$ is
produced by the different nature of the phase transitions.

\section{Comments on uncertainties in the large sub-volume extrapolation}
\label{sec:sys-error}

The sub-volume method has a subtlety in the extrapolation of the
sub-volume data to the large volume limit because the data with
$l\simge N_S/2$ can not be used in the extrapolation.
Thus, special care has to be paid to the estimate of the systematic
uncertainty associated with the subvolume extrapolations.
As described above, we have tried to see the systematic uncertainty in
several different ways by 1) performing the linear and quadratic
extrapolations, 2) comparing the results from different $N_S$ and 3)
comparing with the full volume results if available.
In order to further explore the uncertainty, 4) the fit range in
the extrapolation is varied.
We shift the fit range towards larger $l$ (to be explicit, from
$l=[8,16]$ to $l=[12,20]$) and repeat the same analysis.
Since we would like to keep $l< N_S/2$ to avoid the boundary effects,
the analysis with the new fit range is possible only on $48^3\times 8$
lattices.
%%%%%%%%%%%%%%%%%%%%%%%%%%%%%%%%%%%%%%%%
\begin{figure}[tb]
 \begin{center}
  \begin{tabular}{cc}
   \includegraphics[width=0.5 \textwidth, trim=30 0 0 0]{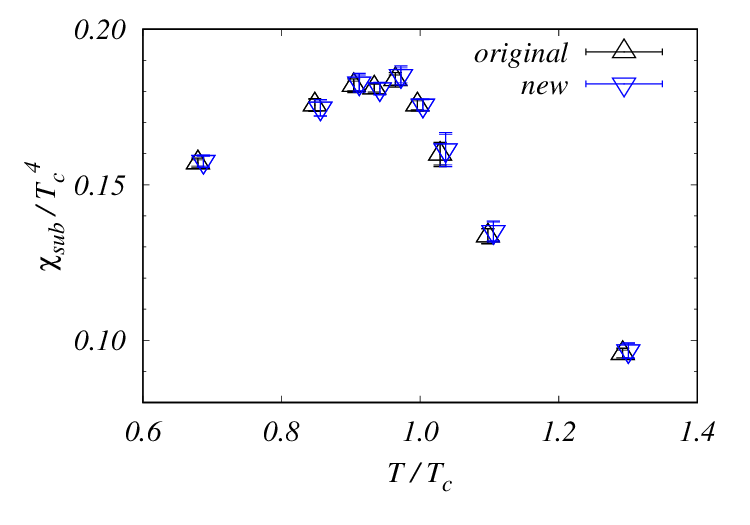}&
   \includegraphics[width=0.5 \textwidth, trim=30 0 0 0]{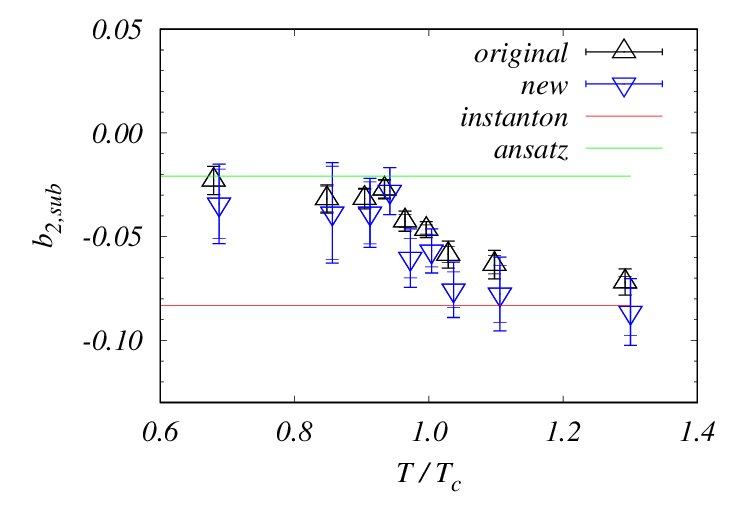}
  \end{tabular}
 \end{center}
 \caption{$T$ dependence of $\chi/T_c^4$ (left) and $b_2$ (right) from
 the sub-volume  method on $48^3\times 8$.
 The results with the original and new fit ranges are compared.
 }
 \label{fig:bdep-chi-b2-2}
\end{figure}
%%%%%%%%%%%%%%%%%%%%%%%%%%%%%%%%%%%%%%%%

Figure~\ref{fig:bdep-chi-b2-2} shows the comparison of the result with
the original and new fit ranges both obtained with $N_S=48$.
For $\chi$, no difference is observed between two fit ranges, and the
estimate of the systematic error is validated.
It is seen for $b_2$ that, while the two fit ranges result in
consistent values and the $T$ dependence, the large statistical errors in
the new results make the difference caused by the different fit ranges
ambiguous.
Although our calculation does not aim at the accurate determination of
$b_2$, the conservative estimate of the error should take into
consideration the spread observed in Figure~\ref{fig:bdep-chi-b2-2}.

\section{\texorpdfstring{$b_2$}{b(2)} in the confined phase}
\label{sec:discussion}

% kokomade
In the previous section, we have seen that $b_2$ is almost $T$-independent
separately in the confined and deconfined phases.
While in the deconfined phase the observed asymptotic value $b_2=-1/12$
is expected from the instanton calculus and asymptotic freedom, there is no obvious explanation for the value in the 
confined phase.
Below we speculate on the reason for the observed constant value of $b_2$
in the low $T$ phase.

\begin{figure}[tb]
 \begin{center}
  \begin{tabular}{c}
   \includegraphics[width=0.7 \textwidth, trim=30 0 0 0]{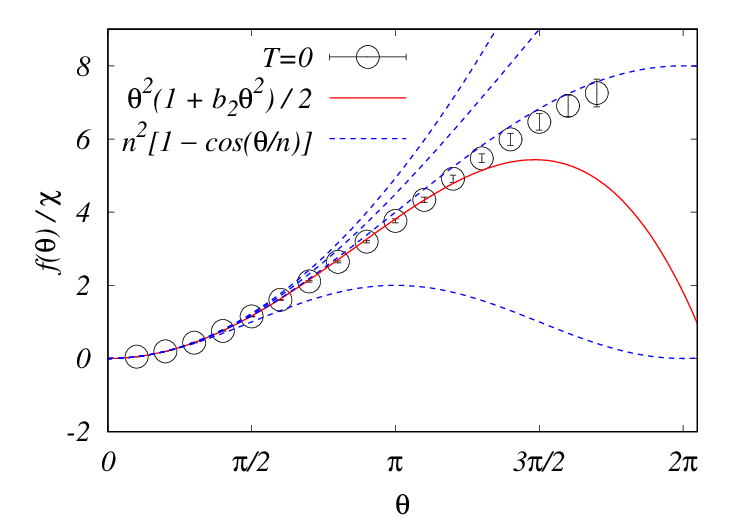}
  \end{tabular}
 \end{center}
 \caption{The $\th$-dependence of the free energy density at
 $T=0$~\cite{Kitano:2021jho}.
 The curves shown are $\th^2(1+b_2\th^2)/2$ with $b_2=-0.023$ (solid)
 and $n^2(1-\cos(\th/n))$ with $n=1,2,3, \infty$ from bottom to top
 (dashed).
 }
 \label{fig:th-f-zeroT}
\end{figure}

Assuming that $b_2=-0.023(7)$ obtained at our lowest temperature
is valid even down to $T=0$, we compare the curve
$f(\th)/\chi=\th^2(1+b_2\th^2)/2$ with the numerical results for the
vacuum energy density calculated in~\cite{Kitano:2021jho}.
In Fig.~\ref{fig:th-f-zeroT}, one finds that the curve well describes the
numerical results up to around $\th\sim \pi$.
To eliminate the discrepancy for $\th > \pi$, we further assume that a
branch of the vacuum energy density $f(\th)$ has a periodicity of
$2n\pi$ in $\th$, where $n$ is an unknown natural number.
One of the simplest possibilities is $f(\th)/\chi=n^2(1-\cos(\th/n))$,
where the overall factor is chosen such that its Taylor expansion
reproduces \eqref{eq:Taylor-e}.
$n=1$ corresponds to the instanton prediction realized in the deconfined
phase.
To find the best value of $n$, $n^2(1-\cos(\th/n))$ is drawn for
$n=1, 2, 3, \infty$ in the same plot.
%%%%%%%%%%%%%%%%%%%%%%%%%%%%%%%%%%%%%%%%
%%%%%%%%%%%%%%%%%%%%%%%%%%%%%%%%%%%%%%%%
We find that the discrepancy is minimum at $n=2$, which
corresponds to $b_2=-1/48\approx 0.021$.

In the plot of $b_{2,\rm sub}$ in Fig.~\ref{fig:bdep-chi-b2}, a
horizontal line has been drawn at $b_2=-1/48$.
The agreement of the data and $-1/48$ is remarkable below $0.95\, T_c$.
From this observation, it is natural to 
deduce that the $\th$-dependence
of the free energy density, normalized by the topological
susceptibility, is approximately $4(1-\cos(\th/2))$ in the confined phase and
$1-\cos\th$ in the deconfined phase at least in the vicinity of
$\th=0$.

In~\cite{Kitano:2020mfk}, $b_2$ of \SU(2) YM is calculated at $T=0$ in
the full-volume method, and collecting the results of $b_2$ from other
\SU($N_c$) theories~\cite{DelDebbio:2002xa,Bonati:2016tvi}, the authors
tested whether the $N_c=2$ theory is large $N_c$-like or instanton-like,
{\it i.e.} $b_2\sim 1/N_c^2$ or $-1/12$.
While the conclusion was against the instanton-like, it was difficult to 
numerically verify that the large-$N_c$ prediction is valid down to $N_c=2$
due to the large statistical error of $b_2$ at $N_c=2$.
In order for the large $N_c$ theory to accommodate $b_2=-1/48$ at
$N_c=2$, there has to be a sizable next-order corrections of $O(1/N_c^4)$;
another possibility is that $b_2$ is a discontinuous function of $1/N_c$
between $N_c=2$ and 3.
The improved determination of $b_2$ will be of help in further
elucidating the consistency with the large $N_c$ predictions.

\section{Summary}
\label{sec:summary}

The sub-volume method was originally developed to estimate the
$\th$-dependence of the free energy density to $\theta\sim O(1)$
without any expansion in~\cite{Kitano:2021jho}.
In this paper this method is applied to the calculation of the topological
susceptibility $\chi$ and the fourth cumulant $b_2$ around the critical
temperature $T_c$.
We have successfully reduced the statistical uncertainties of
both quantities compared to those in the standard full-volume method.
Thanks to the improvement, it is found that $\chi$ has a peak around
$T_c$, and that $b_2$ behaves approximately as a constant both inside
the confined and deconfined phases---$b_2$ in the deconfined phase is
close to the instanton prediction, $-1/12$, as expected, while the one
in the confined phase turns out to be consistent with $-1/48$ down to
$T=0$.
From this observation, we deduced that $\th$-dependence of
the free energy density, normalized by the topological susceptibility,
in \SU(2) YM theory is close to $4(1-\cos(\th/2))$ in the confined phase
at least in the vicinity of $\th=0$.
This proposal is consistent with the expectation that there are two
meta-stable branches of the \SU(2) theory, each of which has $4\pi$
periodicity~\cite{Yamazaki:2017ulc,Nomura:2017zqj}.

We are currently investigating the whole $\th$-dependence of the free
energy density $f(\th)$ at finite temperature.
In order to understand the whole $\th$-$T$ phase diagram, it is also
important to study the $\th$-dependence of the transition temperature $T_c(\th)$
and what happens to $f(\th)$ when it crosses the transition line.

In our calculation with the sub-volume method, the large-sub-volume
extrapolation has been performed in several different ways to estimate
the systematic uncertainties.
There is room for reducing the uncertainty by increasing the data points,
which requires larger lattices.
It is also interesting to explore the relationship between our low-$T$
value $b_2=-1/48$ at $N_c=2$ and the large $N_c$ scaling $b_2\sim
1/N_c^2$ valid, at least, down to $N_c=3$.
We will leave those tasks to future works.

%%%%%%%%%%%%%%%%%%%%%%%%%%%%%%%%%%%%%%%%%%%%%%%
\section*{Acknowledgments}

This work is based in part on the Bridge++ code~\cite{Ueda:2014rya}
and is supported in part by JSPS KAKENHI Grant-in-Aid for Scientific
Research (Nos.~19H00689 [RK, NY, MY], 21H01086~[RK], 22K21350~[RK],
22K03645 [NY], 19K03820, 20H05860, 23H01168 [MY]), and JST, Japan
(PRESTO Grant No.~JPMJPR225A, Moonshot R\&~D Grant No.~JPMJMS2061 [MY]).
This research used computational resources of Cygnus and
Wisteria/BDEC-01 Odyssey (the University of Tokyo), provided by the
Multidisciplinary Cooperative Research Program in the Center for
Computational Sciences, University of Tsukuba.

%%%%%%%%%%%%%%%%%%%%%%%%%%%%%%%%%%%%%%%%%%%%%%%

\end{document}